\documentclass[fleqn
]{an}
\usepackage{graphicx}
\usepackage{times}
\begin{document}

\Pagespan{100}{}
\Yearpublication{2009}%
\Yearsubmission{2008}%
\Month{9}%
\Volume{330}%
\Issue{1}%
 \DOI{10.1002/asna.200811132}%

\title{Edge-on disk galaxies in the SDSS DR6: Fractions of bulgeless and other disk galaxies}

\author{S.J. Kautsch\inst{}\thanks{Corresponding author:
{stefan@astro.ufl.edu}}
}
\titlerunning{Fractions of bulgeless and other disk galaxies}
\authorrunning{S.J. Kautsch}
\institute{Department of Astronomy, University of Florida, 211 Bryant Space 
             Science Center, Gainesville, FL 32611-2055, USA
}

\received{2008 Sep 29}
\accepted{2008 Nov 17}
\publonline{2008 Dec 28}


\abstract{%
The aim of this study is to determine the fractions of different spiral galaxy types, especially bulgeless disks, from a complete and homogeneous 
sample of 15\,127 edge-on disk galaxies extracted from the sixth data release from the Sloan Digital Sky Survey. The sample is divided in broad morphological 
classes and sub types consisting of galaxies with bulges, intermediate types and galaxies which appear bulgeless. A small fraction of disky 
irregulars is also detected. The morphological separation is based on automated classification criteria which resemble the bulge sizes and the 
flatness of the disks. 
Each of these broad classes contains about 1/3 of the total sample. Using strict criteria for selecting pure bulgeless galaxies leads to a fraction of 15\% 
of simple disk galaxies. We compare this fraction to other galaxy catalogs and find an excellent agreement of the observed frequency of bulgeless galaxies. 
Although the fraction of simple disk galaxies in this study does not represent a ``cosmic'' fraction of bulgeless galaxies, it shows that the relative abundance of pure disks is comparable to other studies 
and offers a profound value of the frequency of simple disks in the local Universe. This fraction of simple disks emphasizes the challenge for formation and evolution models of 
   disk galaxies since these models are hard pressed to explain the observed frequency of these objects.
}

\maketitle

\section{Introduction}
Simple disk galaxies are flat, late-type disk galaxies of the morphological Hubble class
$\sim$Sd and later without a bulge component (e.g., Goad \& Roberts 1979, 1980; Karachentsev 1989; Karachentsev et al. 1992; Kautsch et al. 2005b). 
The formation and evolution of such thin galaxies is not yet well understood in the framework 
of $\Lambda$ cold dark matter models. 
Simulations have difficulties in producing 
disk-dominated and bulgeless galaxies. 
The simulated disks are smaller, denser and have lower 
angular momentum than observed, known as the angular momentum problem. Adding feedback processes to the simulations can improve the creation of disk galaxies 
to some extent (e.g., Okamoto et al. 2005; Scannapieco et al. 2008), but D'Onghia \& Burkert (2004), D'Onghia et al. (2006),
K\"{o}ckert \& Steinmetz (2007) and Piontek \& Steinmetz (2008) show that producing the observed structural and kinematical 
properties of disk-dominated galaxies and simple disks cannot be improved by adding feedback processes and by increasing the numerical resolution of the simulations. 
Therefore, the formation mechanisms of simple disks remain enigmatic.

During and after their formation, disk-dominated galaxies and simple disks are very sensitive to various processes that are responsible for transforming those objects into bulge-dominated
galaxies or even destroy the disks. 
The majority of disk-dominated and simple disk galaxies are located in group and low-density environment (Kautsch, Gallagher \& Grebel 2008). Hence, 
this type of environment should preserve the bulgeless shape of those
galaxies. The contrary is observed as mergers are the dominant interaction process in groups (Barnes 1985) and the group 
environment is effective in transforming morphologies (Kautsch et al. 2008b; Tran et al. 2008). Minor mergers already lead to bulge growth 
(e.g., D'Onghia et al. 2006; Cox et al. 2008b; Kazantzidis et al. 2008) and the heating of the thin, stellar disk (Purcell, Kazantzidis \& Bullock 2008). 
Near equal mass mergers can take
place across all different galaxy environments where disk-dominated and simple disk galaxies reside (Karachentsev, Karachentseva \& Parnovskij 1993;
Kautsch, Grebel \& Gallagher 2005) 
and affect the overwhelming majority of Milky Way-sized halos as shown 
in merger-tree statistics within $\Lambda$CDM N-body simulations (Stewart et al.
2008). It is assumed that $\sim$70\% of these Milky Way-sized 
halos contain disk-dominated galaxies and $\sim$11\% of the halos are the hosts of simple disks (Stewart et al. 2007, and referenced therein).
Undergoing such a merger event is basically {\it infaust} for a disk-dominated galaxy and especially simple disks, 
i.e., N-body simulations of mergers exhibit that the stellar
disk can be -- but is not always (Koda, Milosavljevic \& Shapiro 2007; Hopkins
et al. 2008) -- completely disrupted and morphologically transformed into an early-type galaxy
(e.g., Toomre 1977; Steinmetz 2003; Cox \& Loeb 2008).

So called ``pseudobulges'' can grow due to internal disk instabilities. In this secular evolution model gas 
sinks into the disk center and the stars from a subsequent central star-formation period form a bulge (Kormendy \& Kennicutt 2004; Kormendy \&
Fisher 2005).
Galactic bars support the gas flow towards the galactic centers and thus are important for secular evolution. Bars are frequently detected 
in bulgeless galaxies (Barazza, Jogee \& Marinova 2008), making simple disk galaxies potential candidates for ongoing secular evolution. 

The first comprehensive catalog of edge-on disk-dominated galaxies is the ``Flat Galaxy Catalog'' 
(FGC, Karachentsev et al. 1993) and its extension, the ``Revised Flat Galaxy Catalog'' 
(RFGC, Karachentsev et al. 1999). FGC and RFGC are optical all-sky surveys. RFGC contains 4236 
visually selected ``flat'' galaxies. A collection of disk-dominated galaxies in the near-infrared is gathered 
in ``The 2MASS-selected Flat Galaxy Catalog'' (Mitronova et al. 2004).

In order to contribute to a better characterization of flat galaxies, Kautsch et al. (2005b; 2006a) carried out 
a work which compiled a uniform sample of disk-dominated galaxies in the optical wavelengths from the Sloan Digital Sky 
Survey (SDSS, York et al. 2000). 
The SDSS is ideal for the identification 
of such galaxies with its deep, multi-wavelength, homogeneous and large-area coverage.
They analyzed SDSS data from the Data Release 1 (DR1, Abazajian et al. 2003) and compiled a catalog 
of 3169 galaxies with prominent edge-on stellar disks 
(Kautsch et al. 2006a, hereafter ``the catalog,'' which is accessible on-line\footnote{{http://vizier.cfa.harvard.edu/viz-bin/VizieR?-source=J/A+A/445/765}} and Kautsch et al. 2006b). 
Using an automated algorithm, galaxies in the catalog are divided into galaxies with bulges, intermediate types and simple disk galaxies and subclasses. 
15.8\% of the catalog galaxies are found to be simple disks. This demonstrates that bulgeless galaxies are frequent, especially among intermediate-mass 
star-forming galaxies (Matthews \& Gallagher 1997). This frequency increases up to 1/3 of the catalog galaxies when puffy disks are included. The boundaries between the galaxy types in the catalog
are not sharp, suggesting simple disks are the faint end in a continuum of disk galaxy properties, e.g., surface brightness. 

Since DR1, SDSS has undergone significant changes. DR1 provides a survey area of 2099 deg$^2$ 
of imaging data. The data releases DR2 (Abazajian et al. 2004), DR3 (Abazajian et al. 2005), DR4 (Adelman-McCarthy et al. 2006), 
DR5 (Adelman-McCarthy et al. 2007) and DR6 (Adelman-McCarthy et al. 2008) are now available. DR6 covers 9583 deg$^2$ of imaging in total and 
has an $r$-band depth of approximately 22.2 mag. Hence, DR6 contains more than four times the footprint area of DR1. 
In addition, changes were made between DR1 and DR2 concerning the deblending of overlapping objects. 
This correction should improve the detection of individual galaxies.

In this study we collect edge-on disk galaxies as done in the catalog (Kautsch et al. 2006a) but use the larger and newer 
database of the SDSS DR6 and we apply improved morphological separation criteria. 
This allows us to study the fractions of different edge-on disks in a more robust statistical manner. 
The galaxies in this new sample are also divided into morphological classes using a new approach compared to the catalog. 
The improved object detection and especially the much larger coverage area in the sky should provide 
updated statistics, accuracy and homogeneity of the fractions of disk-dominated galaxies in the local Universe. 
In addition, we provide a unique comparison of fractions of bulgeless disk galaxies from several catalogs and derive a robust estimation 
of the frequency of simple disks among spirals in the local Universe. The knowledge of this fraction is crucial for studies about the formation, 
evolution and survival of disk-dominated galaxies. Therefore, this work delivers useful statistical results for several follow-up studies such as the 
fraction of simple disk galaxies at high redshifts. 

This article has the following structure: in Sect. 2 we describe how edge-on disk galaxies were selected in previous works and in this present study. 
The quantified morphological classification of 
galaxies with and without bulges
is discussed in Sect. 3. The results of this study in terms of numbers and fractions 
and a comparison to the fractions of recently published studies are shown in Sect. 4. 
In Sect. 5 we summarize the results.

\section{Data acquisition}

In the catalog, edge-on disk galaxies were selected based on selection criteria that
are similar to those of the original approach by Karachentsev et al.'s catalogs FGC and RFGC. The original object
selection by Karachentsev et al. (1993; 1999) is based on the visual identification of galaxies with an
axial ratio $a/b\geq$7 and a major axis diameter of $\ga$40\arcsec ~in the blue band POSS-1
copies and ESO/SERC photographic plates. In order to come as close as possible to these
original values, a training set of RFGC galaxies that were recovered in the SDSS DR1 
was analyzed in the catalog with respect to their
sizes, magnitudes and color distribution. This allowed Kautsch et al. (2006a) to translate the original object
selection from the RFGC into a query that selects all galaxies brighter 
than ${m_g=20}$ from 
the DR1 ``Best Galaxy Table'' with axial ratios $>$3 and a major axis diameter $>$30\arcsec ~within 
certain broad color limits. For
the details of this query, please consult Kautsch et al. (2006a; 2006b). 
Our intention now is to collect edge-on galaxies from the DR6 in the same reproducible
fashion. 
Therefore we apply the SDSS CasJobs on the SDSS Context DR6 to the 
``Galaxy Table''. CasJobs\footnote{{ http://casjobs.sdss.org/CasJobs}} is an online interface that performs queries on 
various SDSS datasets using the ``Structured Query Language'' (SQL) and the ``Galaxy Table'' is the table of the SDSS data archive that contains all parameters 
for objects selected as galaxies in DR6 with the highest quality at the time of the data release. 
The query used in the catalog and in this study selects all galaxies with: {i)} an axial ratio ${a/b > 3}$ ($a$ major axis, $b$ minor axis) in the $g$-band, 
{ii)} an angular major axis diameter $a > 30 \arcsec$ in the $g$-band, {iii)} colors in the range of $-0.5 < g-r < 2$ and $-0.5 < r-i < 2$, 
{iv)} a magnitude limit in the $g$-band $<20$ mag. These selection criteria have the following form in SQL:
\texttt{ \\
SELECT*\\
into mydb.edge\_on\_dr6\_catalog from \\ Galaxy as G\\
WHERE G.petroMag\_g $<$ 20\\
and (G.isoA\_g/G.isoB\_g) $>$ 3\\
and G.isoA\_g $>$ 37.8 pixel {\it this corresponds to \\ an angular radius of 15\arcsec}\\
and (G.dered\_g - G.dered\_r) between -0.5 \\ and 2\\
and (G.dered\_r - G.dered\_i) between -0.5 \\ and 2} \

This query leads to a sample of 27\,308 objects. A visual inspection shows a non-negligible number
of contaminants. The contaminants are mostly stellar refraction spikes and artifacts such as 
satellite/meteor tracks, empty images as well as face-on galaxies where elongated structures 
(e.g., bars, spiral arms) simulate an edge-on disk appearance. In addition, some of the galaxies 
are also affected by ``shredding'' (Abazajian et al. 2004). 
This means that a single object has more than one unique detection and therefore multiple entries in the 
SDSS database. In the catalog, these contaminants were removed manually. Because 
of the large numbers in the present sample, this approach is inefficient and should be quantified. 
We find that the majority of the contaminants are automatically removed by 
excluding the objects flagged
with the following DR6 PhotoFlags\footnote{These flags are associated with every unique object in the database and contain 
important information of the quality of the object: 
{ http://www.sdss.org/dr6/products/catalogs/flags.html.}}: 
\begin{itemize}
\item ``\texttt{edge}'' indicates galaxies truncated on the survey borders; 
\item ``\texttt{saturated}'' indicates saturated pixels; 
\item ``\texttt{notchecked}'' indicates that SDSS deblending may be unreliable; 
\item``\texttt{too\_few\_good\_detections}'' indicates objects with  no good centroid found in all bands; 
\item ``\texttt{petroMagErr}''$>$ 0.3 in $g,\ r,\ i$ indicates the magnitude errors.
\end{itemize}
This leads to a DR6
sample of 15\,176 objects. We exclude also the objects that do not have 
Petrosian radii, \\ ``\texttt{petroR50\_r}'' and ``\texttt{petroR90\_r}.'' 
The final DR6 edge-on galaxy sample contains 15\,127 objects. 

We select various large random samples in order to estimate the
remaining contribution of contaminants in this final sample. We find a rate of 5\% 
of false detections for the whole sample, mainly due to shredded galaxies, elongated HII regions in disks, projected
objects in spiral arms, empty images, projected objects in halos of saturated stars and rarely, 
stellar spikes that are still present in the sample.


\section{Morphological separation}

   \begin{figure}[t]
\vskip-4mm \hskip-6mm
   \resizebox{\hsize}{!}{\includegraphics[angle=-90]{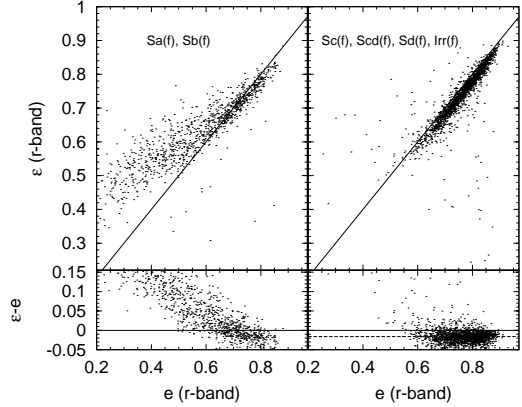}}
      \caption{This figure shows the relation between the luminosity weighted mean ellipticity $\varepsilon$ (ordinate) and the adaptive 
      ellipticity e (abscissa) for all galaxies in the catalog in the SDSS $r$-band. Early-type disk galaxies are plotted in the left part 
      and late-type disks in the right part of this figure. The scatter plots are presented in the lower section of the figure. The fine solid lines 
      in the upper diagrams indicate the
      linear relation with a slope of one. The solid lines in the bottom diagrams show the location of $\varepsilon-{\rm e}=0$. 
      The dashed line in the scatter plot for the late disks (bottom right) indicates
      the derived mean offset between $\varepsilon$ and e.
              }
         \label{e_scatter}
   \end{figure}

In the catalog we used an automated morphological classification in order to define six types of 
edge-on disk galaxies. These are galaxies with bulges (Sa(f), Sb(f)); simple disk galaxies (Sd(f)); 
and disk-dominated intermediate types (Sc(f), Scd(f)) as well as disky edge-on irregulars (Irr(f)). 
We follow the terminology by de Vaucouleurs (1959) where spiral galaxies are flagged with a letter associated with the shape of the 
spiral arms (ring-shaped galaxies have an ``r'' and s-shaped galaxies have an ``s''). We instead 
are using an ``f'' in brackets in order to indicate that the galaxies contain flat disks seen edge on. 

The morphological separation in the catalog is based on the concentration index (CI) and the
luminosity-weighted mean ellipticity of the elliptical isophotes ($\varepsilon$), both in the SDSS $r$-band. 
The CI is a measure of the
bulge size for edge-on galaxies and is defined as the ratio of the Petrosian radii that contain
90\% and 50\% of the Petrosian flux in the same band in a circular aperture. We used
the Petrosian radii that are listed in the ``Galaxy Table.'' 
The CI separation values divide the early (Sa(f), Sb(f)) from the late types (Sc(f), Scd(f), Sd(f)). In addition, low CI values separate
the Irr(f) class from the late types. We use the same CI separation criteria as in the catalog because the Petrosian radii 
did not undergo changes between DR1 and DR6.

$\varepsilon$ is a discriminator of the
flatness of edge-on disks and was measured directly on the individual galaxy images of the galaxies in
the catalog. This
measurement was performed with the MIDAS {surfphot} package. We fit elliptical
isophotes to the galaxies and derived $\varepsilon$ as the weighted mean ellipticity of all isophote
levels of an individual galaxy. 

We replace the method finding $\varepsilon$ by using variables offered directly from the SDSS tables. 
The new method offers consistency within the SDSS parameters. In addition, with this method it is simple to reproduce 
our morphological separation criteria in a fast and efficient way.
Disk flatness can be  
expressed through the adaptive moments provided in the DR6 ``Galaxy Table.'' 
The moments are measured from the ellipticity and size of the objects within the
SDSS
pipeline\footnote{http://www.sdss.org/dr6/algorithms/adaptive.html}.
The adaptive second moments from the SDSS (\texttt{mE1} and \texttt{mE2}),
\begin{eqnarray}
e_+ & =  & \texttt{mE1},\\
e_{\times} &  =  & \texttt{mE2},
\end{eqnarray}
can be converted into an ``adaptive'' ellipticity (e), using $a$ and $b$ as major and minor axis, respectively (c.f., Vincent \& Ryden 2005):
\begin{equation}
{\rm e} = 1 - \frac{b}{a} = 1 - \sqrt{\frac{1 - \sqrt{e_+^2 + e_{\times}^2}}{1 + \sqrt{e_+^2 + e_{\times}^2}}}\,.
\end{equation}

We now compare $\varepsilon$ with e for all galaxies from the catalog. 
The upper part of Fig.~\ref{e_scatter} shows the relation between $\varepsilon$ and e 
for the early-disk types (left) and the late types (right). 
The diagrams on the bottom of Fig.~\ref{e_scatter} show the scatter plots. Early disks (Sa(f), Sb(f)) 
(on the left side of this plot) tend to have a
rounder shape (i.e., lower ellipticity values) of e compared with $\varepsilon$. 
In contrast, late disk types Sc(f), Scd(f), Sd(f)
and Irr(f) (shown in the right part of the figure) are
flatter using e instead of $\varepsilon$. We also see some late-type outliers having 
significantly lower $\varepsilon$ compared to e. A visual inspection of the outliers exhibits that many of these points belong to
misclassified Irr(f) types in the catalog.

\begin{figure}
   \resizebox{\hsize}{!}{\includegraphics[width=6cm]{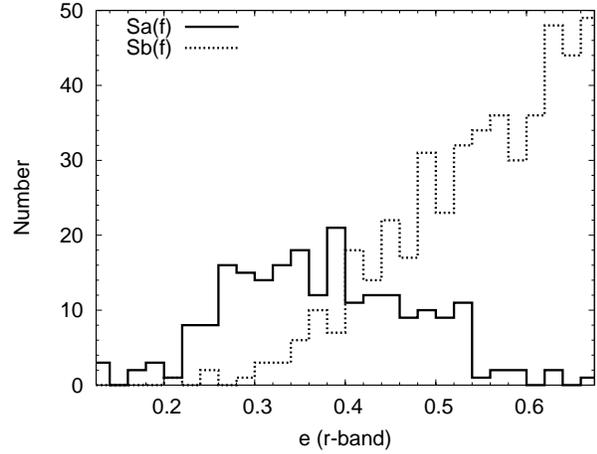}}
      \caption{The number distribution for the early-type disk galaxies of the adaptive ellipticity (e) is shown in this histogram. 
      Sa(f) galaxies, the class containing galaxies with the largest bulges, are drawn with a solid line. Sb(f) galaxies are indicated with a dotted line. 
              }
         \label{hist}
   \end{figure}

The limiting values of $\varepsilon$ in table~1 in the catalog must be recalculated in order to separate the galaxies into different morphologies 
based on e. The late types are constantly shifted from a simple linear relation with a slope of one between $\varepsilon$ and e as shown in the right bottom 
part of Fig.~\ref{e_scatter}. We derive a mean offset of $\varepsilon-{\rm e}=-0.016$, indicated as dashed line in Fig.~\ref{e_scatter}. 
Because the late disks are flatter in e, the offset must be added to $\varepsilon$ in order to resemble the limiting values of the catalog. 
We emphasize that the main intention of this investigation is to study the fractions of disk-dominated galaxies. Therefore, early-type disks (classified as Sa(f)
and Sb(f)) are less important for this purpose. However, we use a histogram of the number distribution of e for the galaxies classified
as Sa(f) and Sb(f) in the catalog. With the aid of this histogram (Fig.~\ref{hist}) we define a value of ${\rm e}=0.4$ as the best separation
value that resembles the dividing limit between Sa(f) and Sb(f).
The
limiting values for CI and e are collected in Table~\ref{tab1}. 

\begin{table}[t]
\caption{The limiting values of the adaptive ellipticity (e) and concentration index (CI) between the morphological classes are listed in this table. 
The values are valid for the SDSS $r$-band.
In general, note that the galaxies
near boundaries have the least certain classification.
}
\label{tab1}
\tabcolsep=2.8pt
\begin{tabular}{lrr@{\hspace{4mm}}rr}
\hline\noalign{\smallskip}
Class &  \multicolumn{2}{c}{e} & \multicolumn{2}{c}{CI}\\[1.5pt]
 \hline\noalign{\smallskip}
  & Lower Limit & Upper Limit &  Lower Limit & Upper Limit\\[1.5pt]
\hline\noalign{\smallskip}
Sa(f) & $ -- $ & $< 0.40 $ & $\geq 2.70 $ & $ -- $ \\
Sb(f) & $\geq 0.40 $ & $ -- $ & $\geq 2.70 $ & $ -- $ \\
Sc(f) & $ -- $ & $< 0.766 $ & $\geq 2.15 $ & $< 2.70 $  \\
Scd(f) & $\geq 0.766 $ & $< 0.816 $ & $\geq 2.15 $ & $< 2.70 $ \\
Sd(f) & $\geq 0.816 $ & $ -- $ & $ -- $ & $< 2.70 $ \\
Irr(f) & $ -- $ & $< 0.816 $ & $ -- $ & $< 2.15 $ \\[1.5pt]
\hline

\end{tabular}
\end{table}

\section{Results}
\subsection{Fractions of different types of edge-on galaxies}

\begin{figure}
   \includegraphics[width=1.6in]{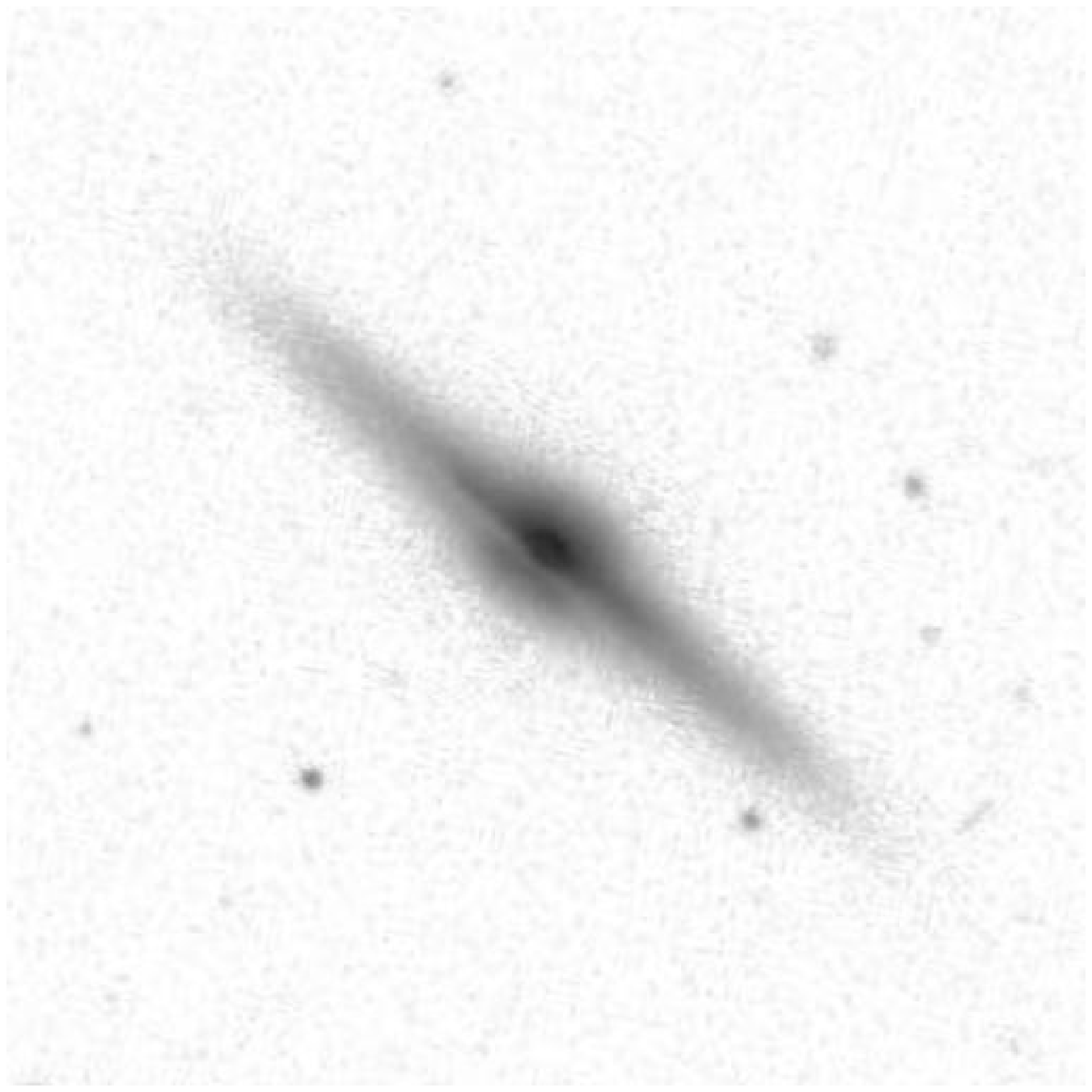}
    \includegraphics[width=1.6in]{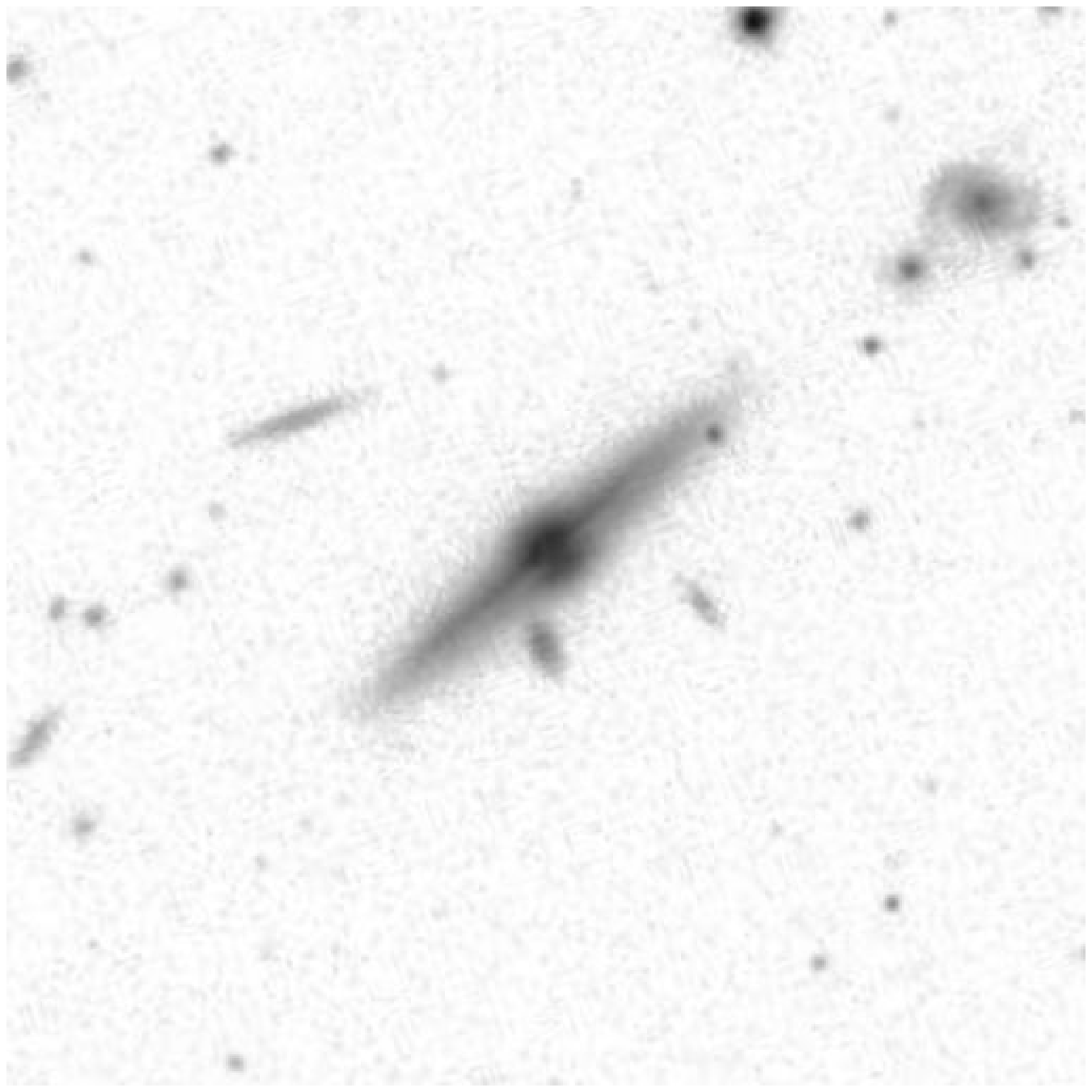} 
      \caption{Sa(f): J111146.36+364442.3 (\emph{right panel}), J143650.68 +294525.5 (\emph{left panel}).}
       \label{sa}
   \end{figure}

\begin{figure}
   \includegraphics[width=1.6in]{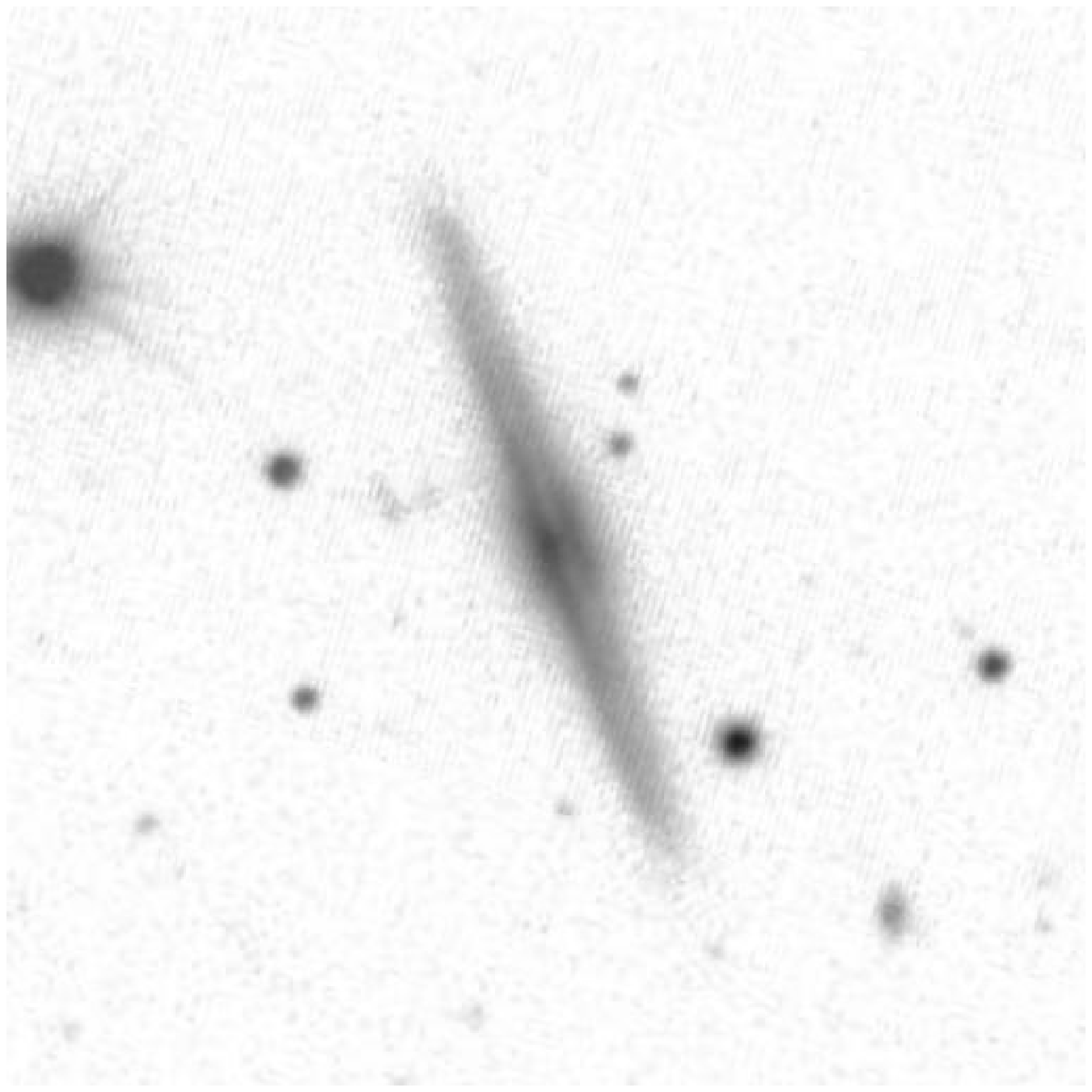}
   \includegraphics[width=1.6in]{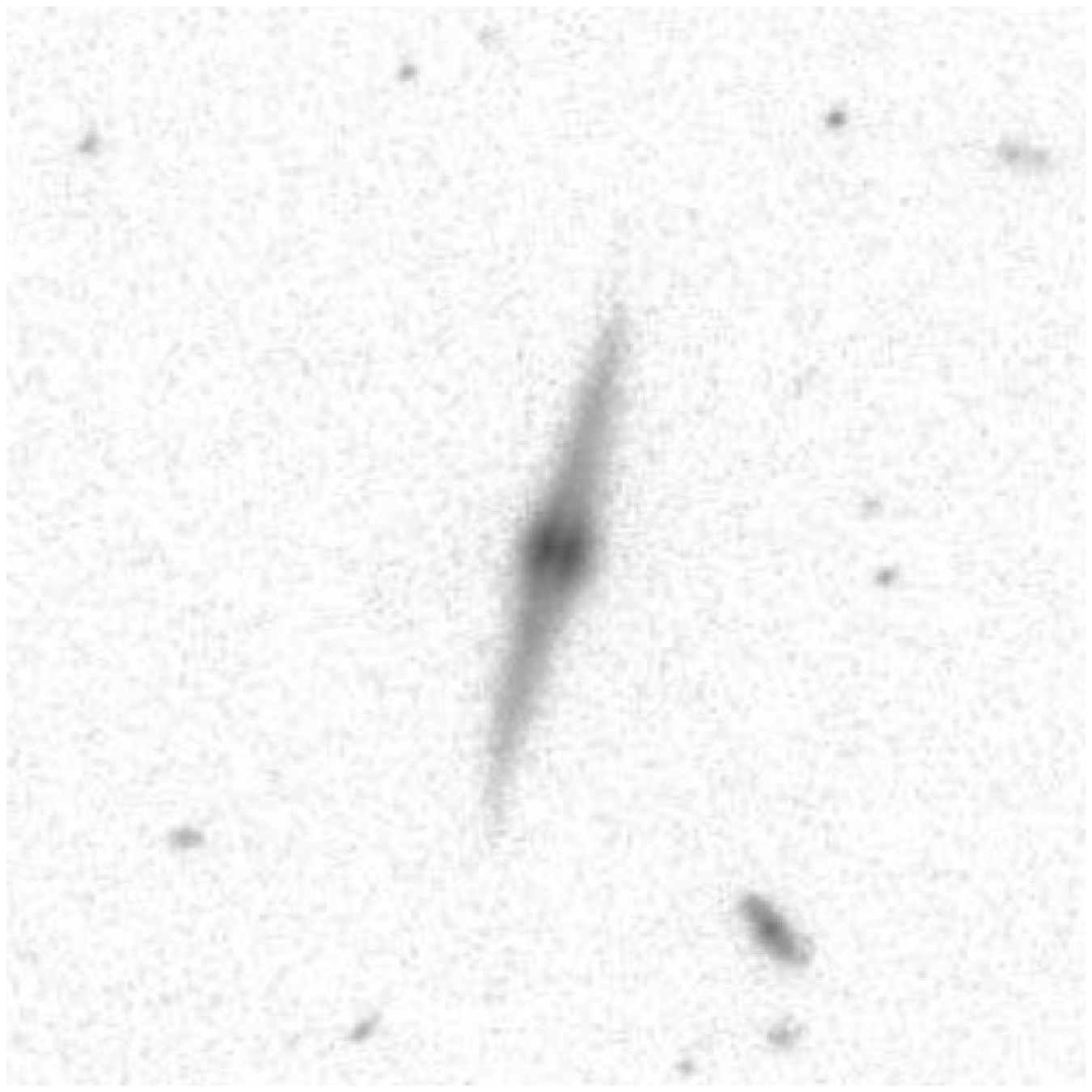} 
      \caption{Sb(f): J172506.84+565241.8 (\emph{right panel}), J231006.72 \hbox{--093953.6} 
              (\emph{left panel}).}
         \label{sb}
   \end{figure}

\begin{figure}
   \includegraphics[width=1.6in]{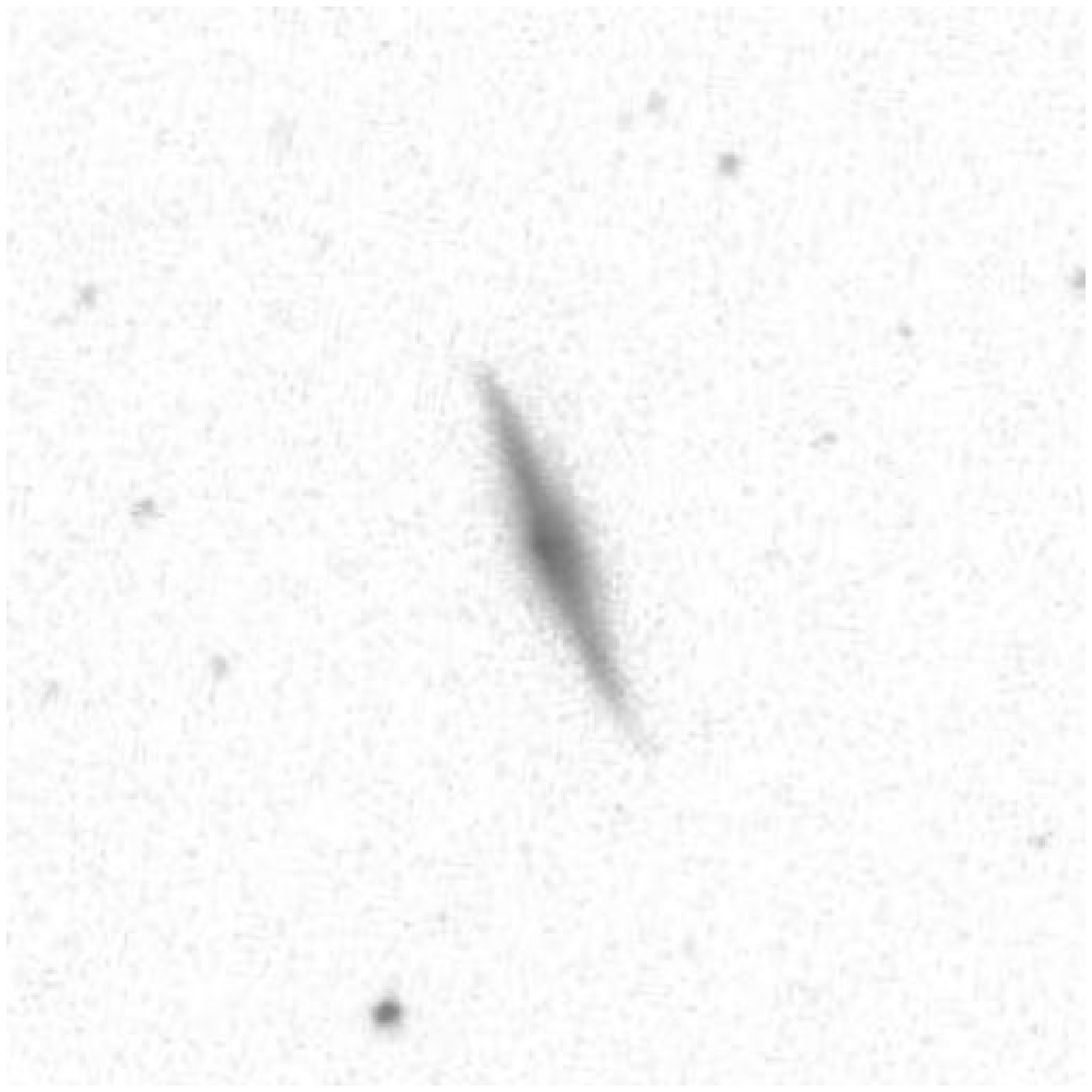}
   \includegraphics[width=1.6in]{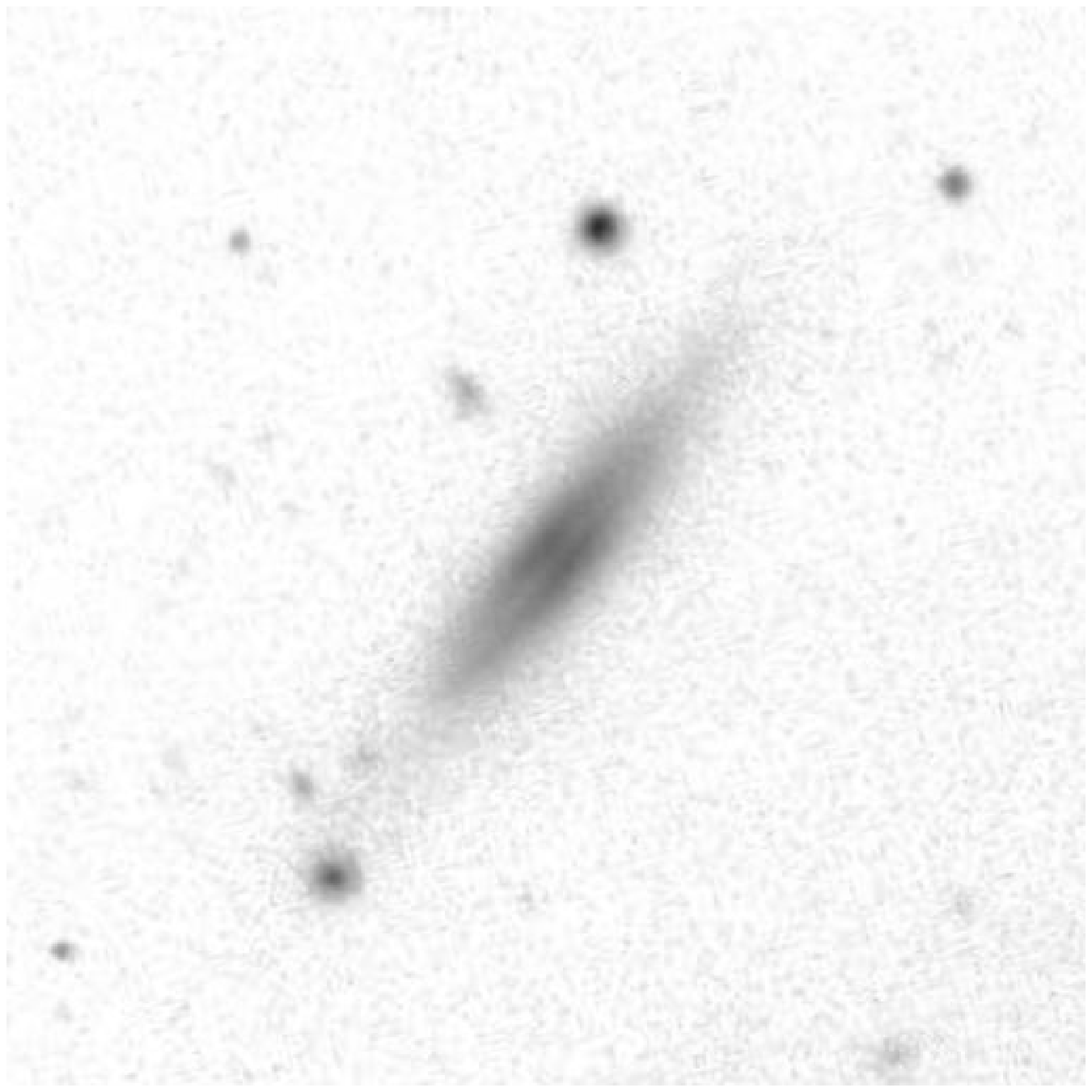}
    
      \caption{Sc(f): J020800.74--082442.1 (\emph{right panel}), J143917.88 +200439.7 (\emph{left panel}).}
         \label{sc}
   \end{figure}

\begin{figure}
   \includegraphics[width=1.6in]{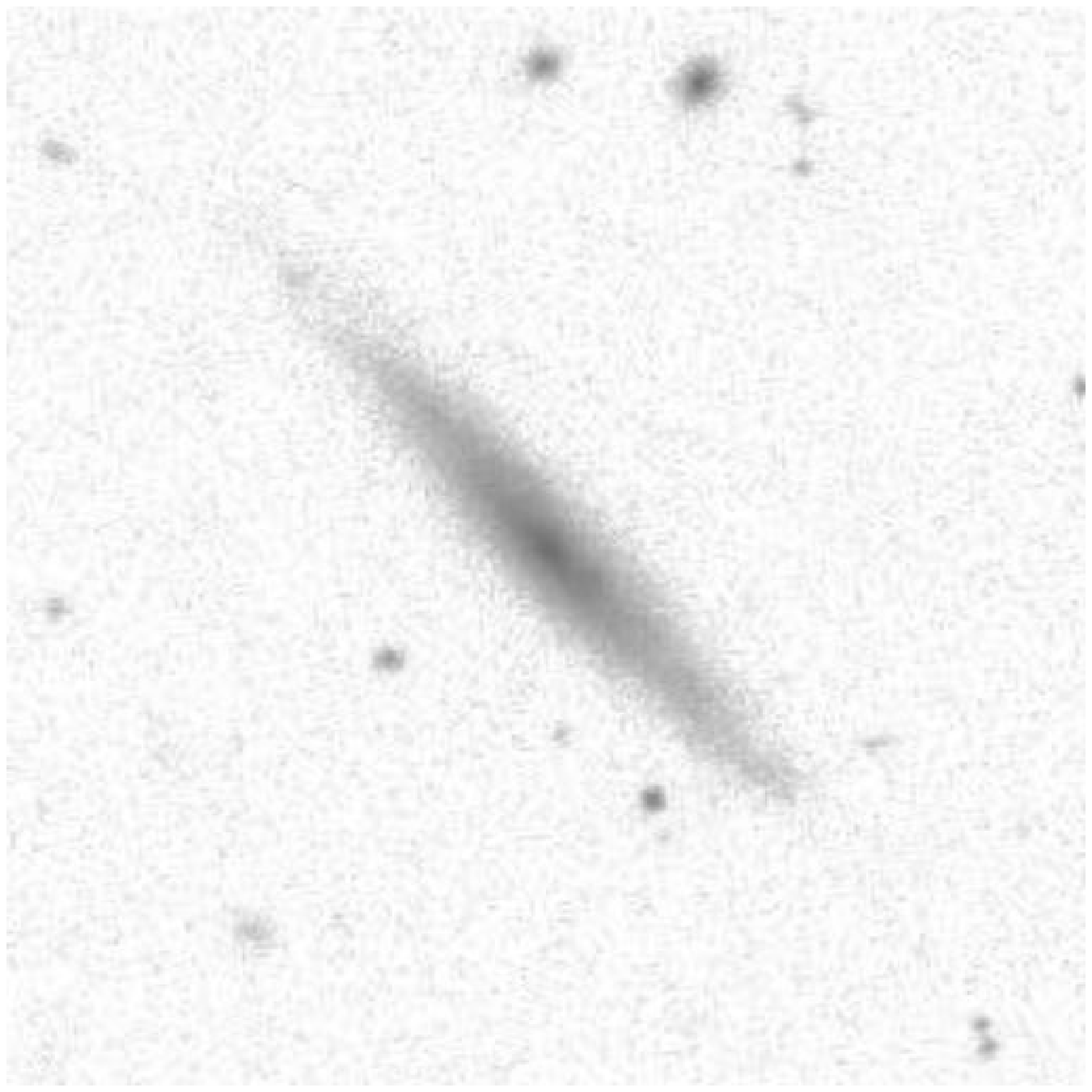}
   \includegraphics[width=1.6in]{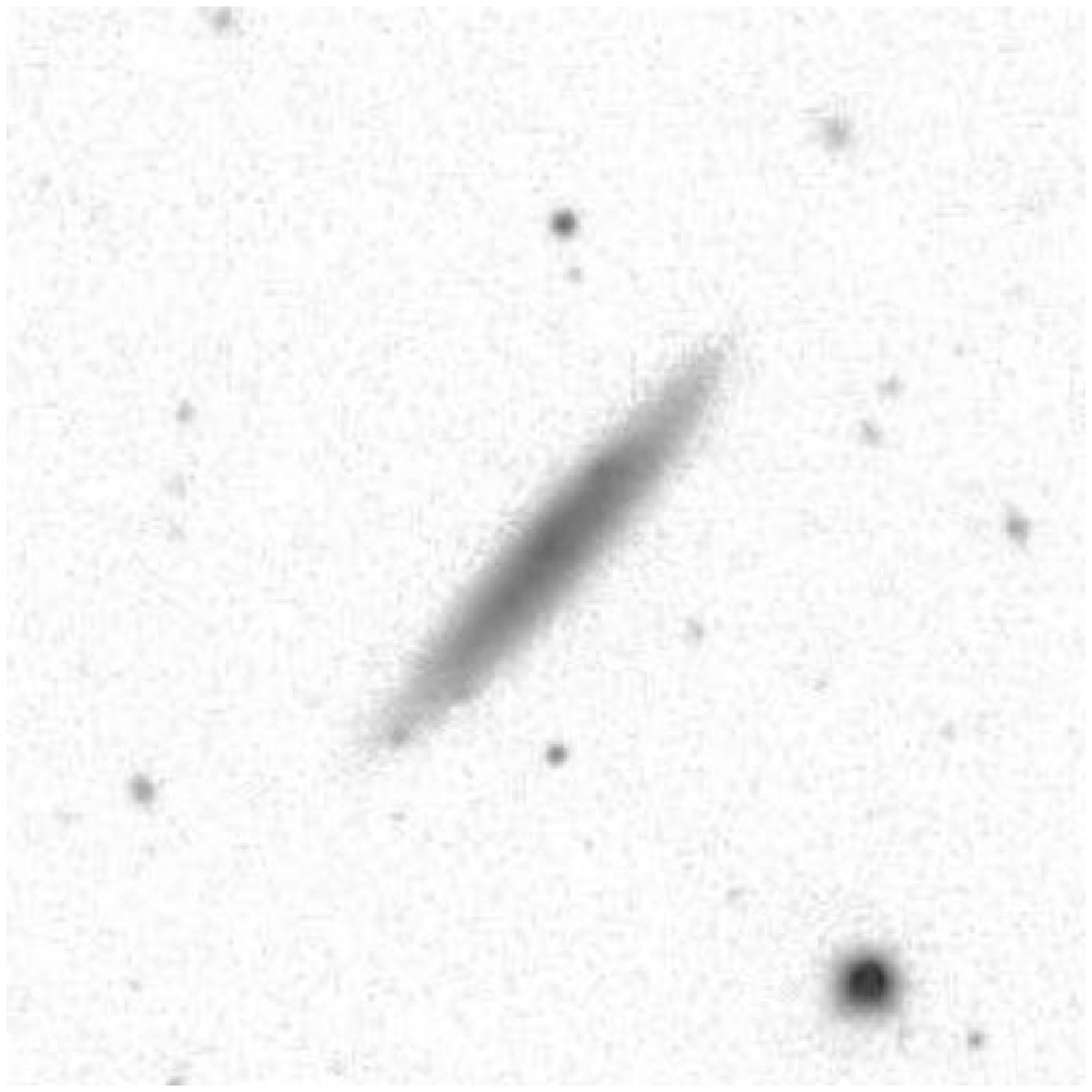}
    
      \caption{Scd(f): J225828.27--103332 (\emph{right panel}), J124943.24 +044610.1 (\emph{left panel}).}
         \label{scd}
  \end{figure}

\begin{figure}
   \includegraphics[width=1.6in]{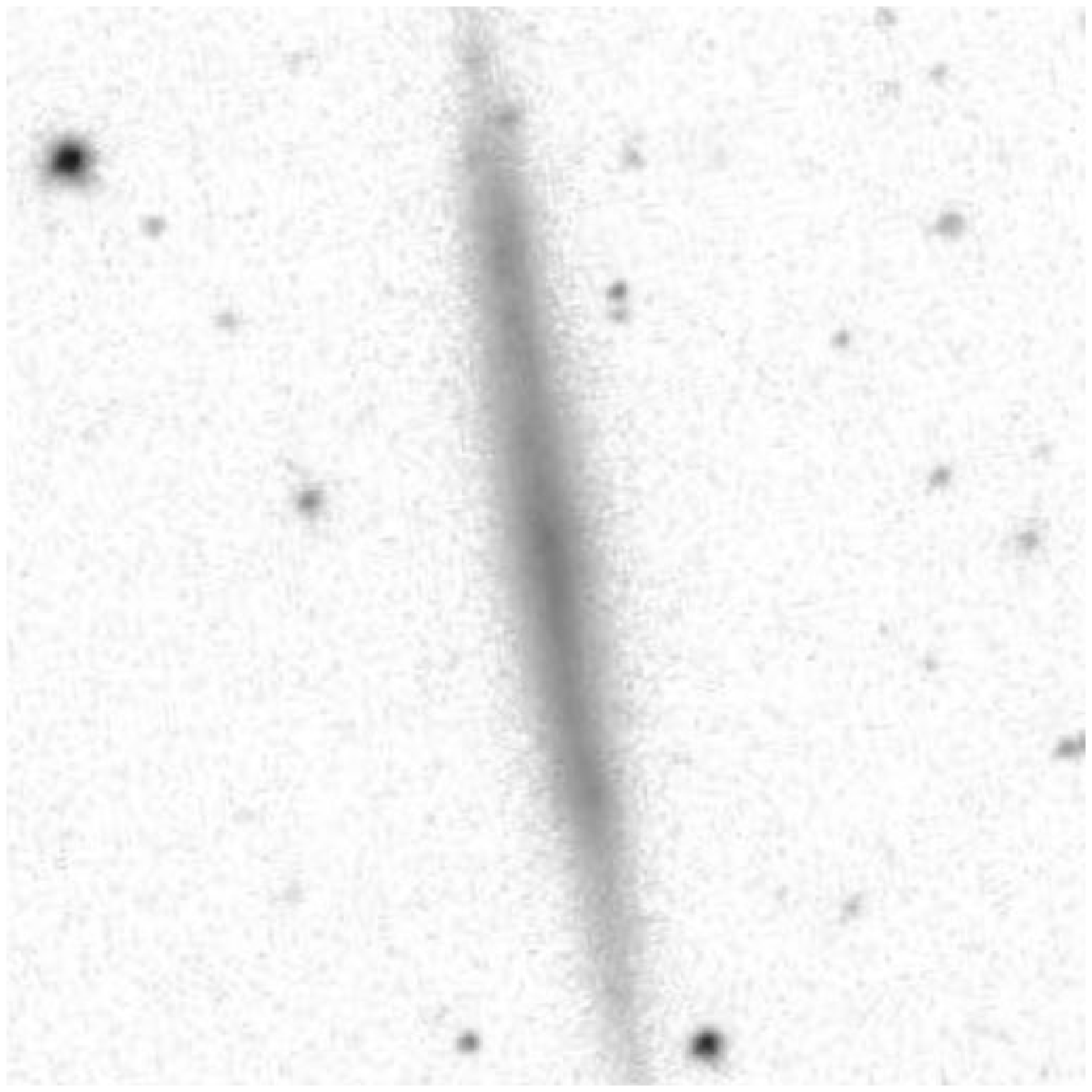}
   \includegraphics[width=1.6in]{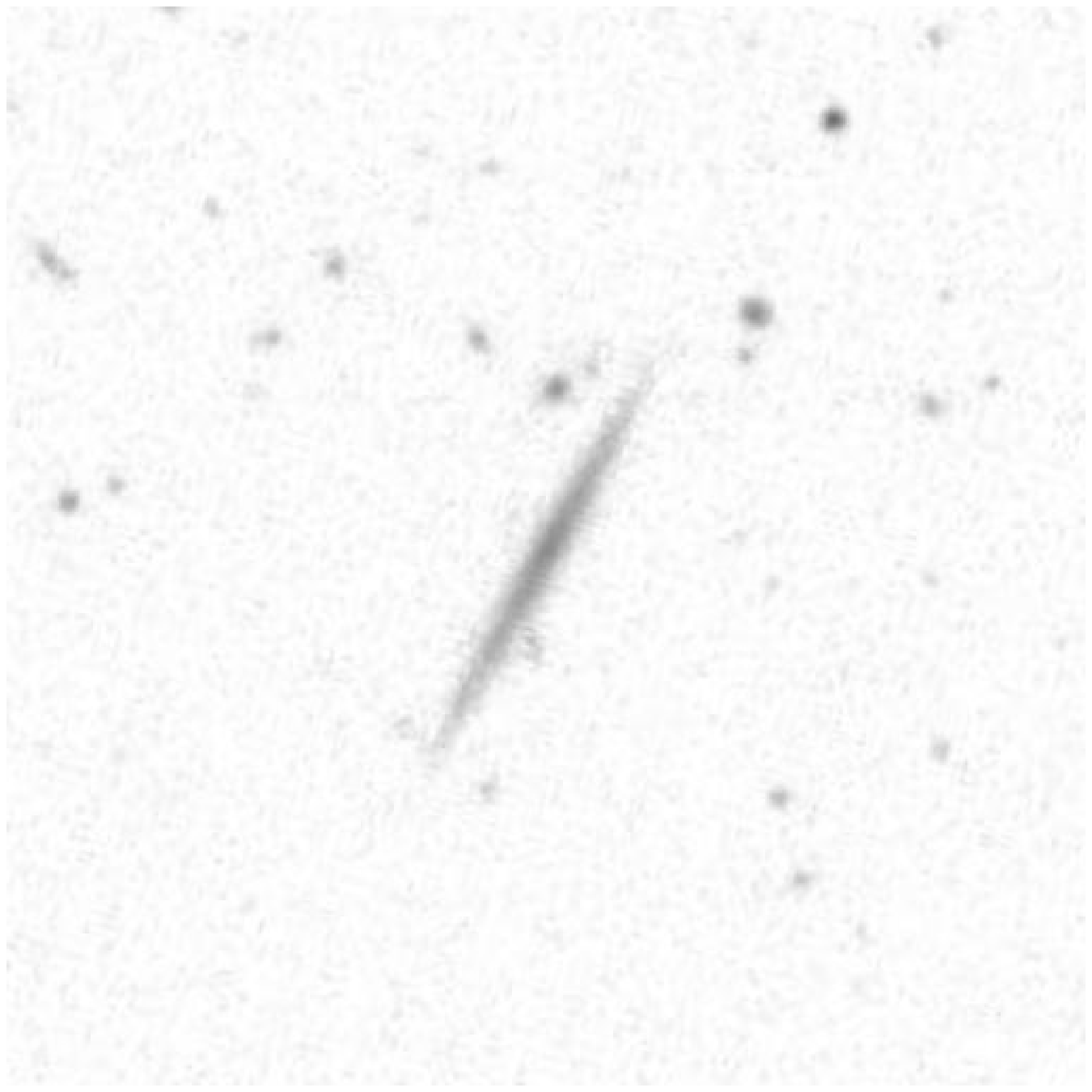}
    
      \caption{Sd(f): J214439.43--064122.5 (\emph{right panel}), J104658.44 +382728 (\emph{left panel}).}
         \label{sd}
   \end{figure}

\begin{figure}
   \includegraphics[width=1.6in]{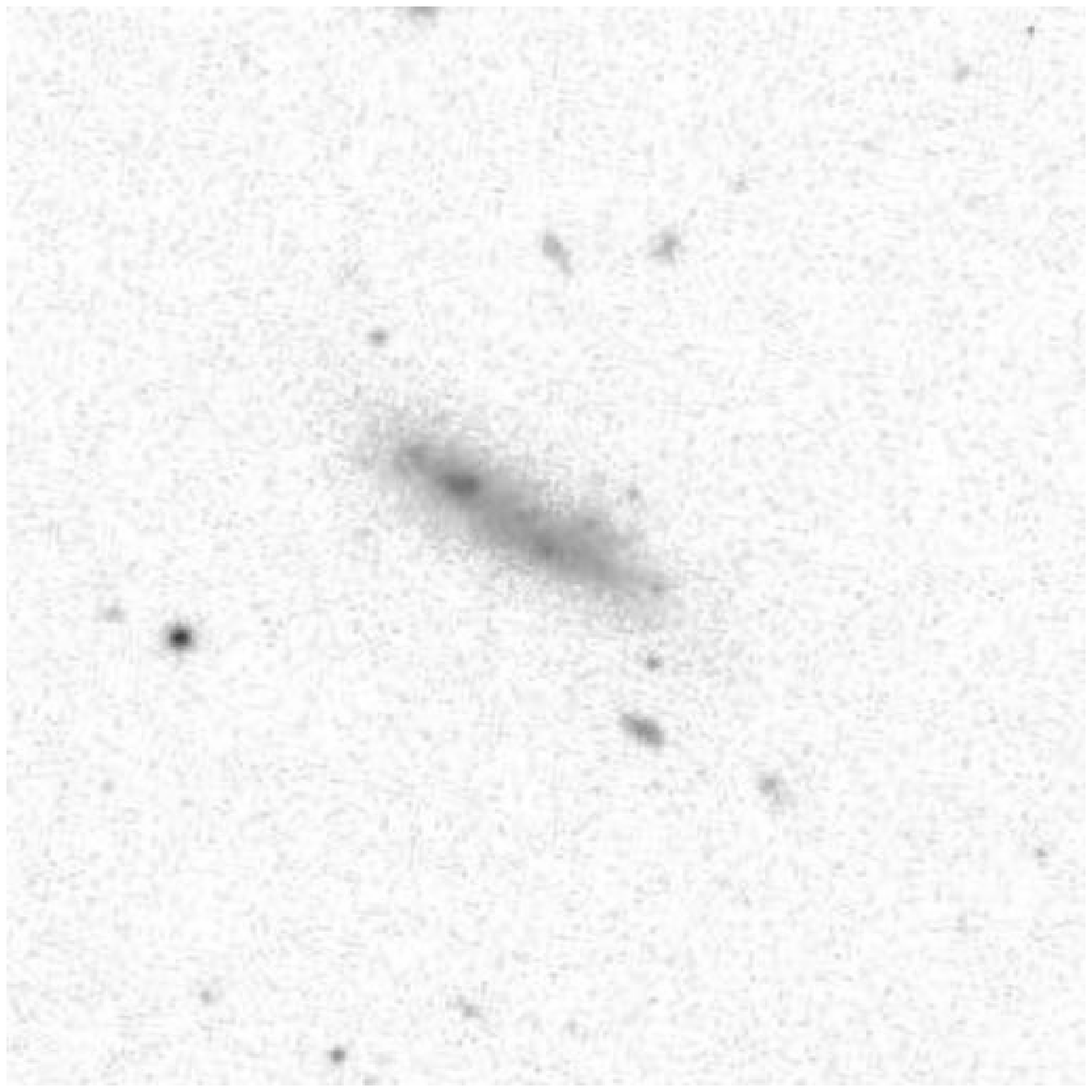}
   \includegraphics[width=1.6in]{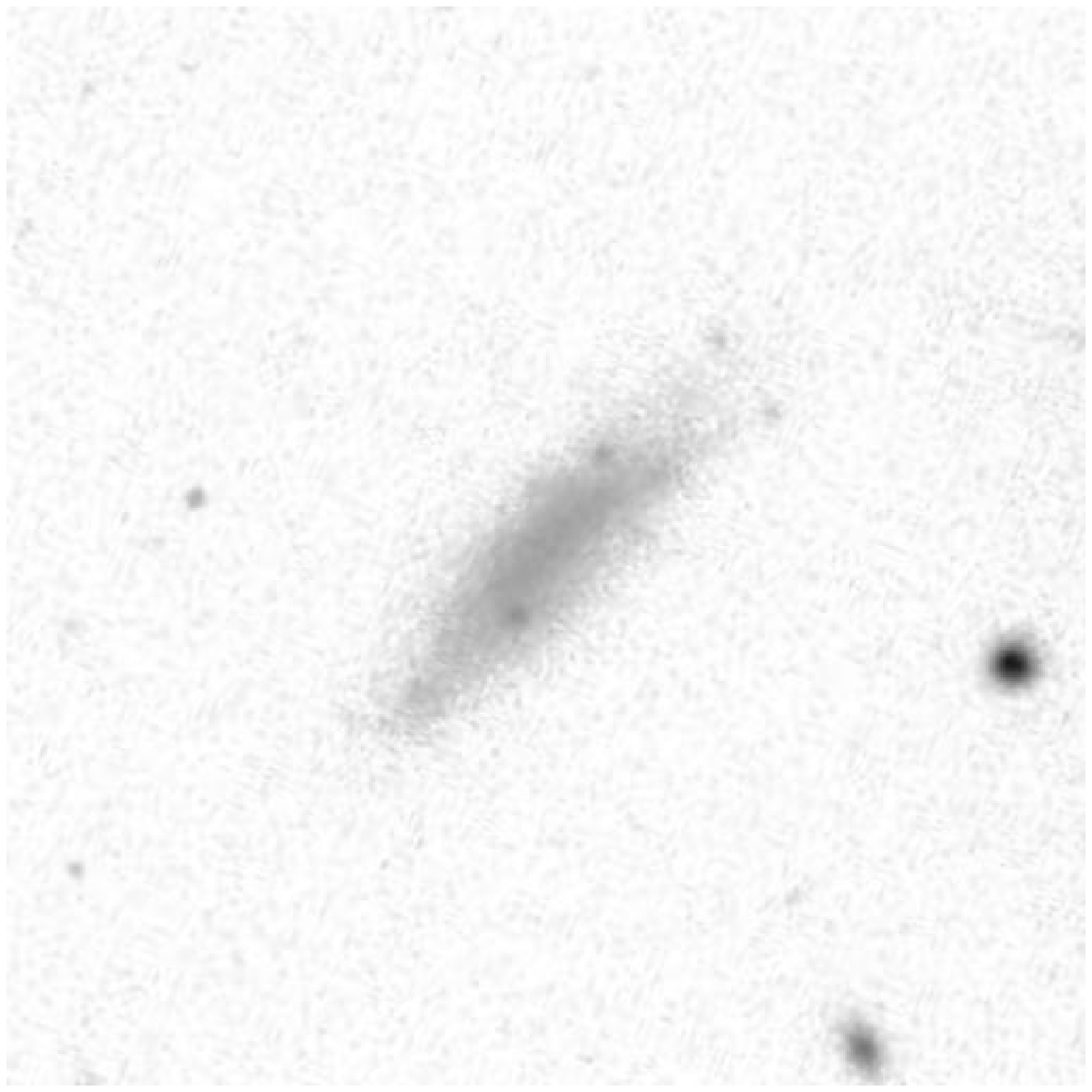} 
      \caption{Irr(f): J091028.24+071111.7 (\emph{right panel}), J104210.63 +632430.5 (\emph{left panel}).}
         \label{irr}
   \end{figure}

We apply the new limiting values from Table~\ref{tab1} to the new DR6 edge-on galaxy sample. 
Figures~\ref{sa},~\ref{sb},~\ref{sc},~\ref{scd},~\ref{sd}, and \ref{irr} show example images of galaxies with types Sa(f), Sb(f), Sc(f), Scd(f), Sd(f) and Irr(f), respectively.
These images have an angular size of 100 square arcsec and north is to the top, east to the left. 
The images are downloaded from the SDSS Image List
Tool\footnote{http://cas.sdss.org/dr6/en/tools/chart/list.asp}.
The number results are shown in Table~\ref{tab2}. In this table we also reprint the
numbers of the classes found in the catalog on the right side. 
The largest difference between the catalog and this study is the fraction of Irr(f). 
Their increased number in the present study is due to
the contamination of misclassified objects, as revealed in a visual inspection of randomly selected Irr(f) types. 
Aside from this, the fractions of the different types are almost the same. 
For completeness considerations of the classes we refer to Sect.~6 in the catalog.

\begin{table}[t]
\caption{The galaxy classes and their fractions are shown in this table. The absolute numbers of
galaxies in the main morphological classes (Col.~2) and their percentages (Col.~3)
are listed in this table.
For comparison the numbers and percentages of the catalog are given in Cols. 4 and 5.}
\label{tab2}
\tabcolsep=3.0pt
\begin{tabular}{l@{\hspace{10pt}}cc@{\hspace{10pt}}cc}
\hline\noalign{\smallskip}
Database &  \multicolumn{2}{c}{SDSS DR6} & \multicolumn{2}{c}{SDSS DR1}\\[1.5pt]
\hline\noalign{\smallskip}
General Class & Number & Percentages & Number & Percentages\\[1.5pt]
\hline\noalign{\smallskip}
Sa(f) &\enspace966 & \enspace 6 & 222 &\enspace 7 \\
Sb(f) & 3993 & 26 & 843 & 26 \\
Sc(f) & 4835 & 32 & 1005\enspace & 32 \\
Scd(f) &2257& 15 & 503 & 16 \\
Sd(f) & 2220 &  15 & 501 & 16 \\
Irr(f) & \enspace856&\enspace6 & \enspace95 &\enspace 3 \\
Total & 15127\enspace & 100\enspace  & 3169\enspace & 100\enspace \\[1.5pt]
\hline
\end{tabular}
\end{table}

The limiting values for Sd(f) of the catalog are based on fairly conservative separation criteria in order to minimize possible
contamination from other classes. So the fraction of simple disks is 15\% in the present study. In the catalog we show that it is also possible to 
include the
Scd(f) types in an extended bulgeless disk class. The fraction of this extended bulgeless disk class (Scd(f) and Sd(f)) is 30\% of 
the total DR6
sample. This value is similar to the value of 32\% of the catalog. Therefore, the fraction of the 
extended bulgeless disk class in this work and the
catalog corresponds to roughly 1/3 of the total sample of edge-on disk galaxies. However, the contamination by other types is larger than with the more
rigorous defined limits for simple disks.

\subsection{Comparison with other studies}

It is generally known that disk galaxies of late Hubble morphologies are often bulgeless (e.g., Hubble 1936; B\"{o}ker et al. 2002; Kormendy \& Kennicutt 2004). 
However, accurate number statistics of the fraction of simple disk galaxies are rare, probably due to the lack of large samples. 

Barazza et al. (2008) studied the fraction of bars in disk galaxies and they also analyzed the fraction of bulgeless galaxies based upon visual inspection. 
The fraction of bulgeless disks is $\sim\! 20\%$ in their sample. The sample they used is selected from the SDSS with ${-18.5\leq M_g<-22}$ and 
redshifts between ${0.01<z<0.03}$. Disk galaxies are selected using a color cut, and galaxies with inclinations larger than 60\degr  ~are omitted. Their fraction of bulgeless galaxies is 
close to that of the simple disks (Sd(f)) found here. The $\sim\! 5\%$ difference can be explained by considering that Barazza et al. (2008) used a color cut for the disk galaxy selection. 
This could exclude red spiral types and therefore slightly offset their sample towards bluer and later disk types. 

Koda et al. (2007) used the ``Tully Galaxy Catalog''\,\footnote{{http://haydenplanetarium.org/universe/duguide/exgg\_tully.php}} in order to select 
galaxies with morphological information given in that catalog. The fraction of Sd galaxies is 11\%. 
Considering the restriction to local (dist $<\! 20\, h^{-1}$, ${-17 < M_B < -20}$) galaxies, their fraction is similar to ours. 
The small difference of ${\sim\! 4}\%$ is probably an effect of visual classification: 
The ``Tully Galaxy Catalog'' is not restricted to edge-on galaxies. Less inclined objects can exhibit central light concentrations from nuclei (e.g., Walcher et al. 2006)
which are not seen edge-on. For this reason, some Sd would have been classified as earlier types. 

In Allen et al. (2006), 10095 galaxies with ${m_B<20}$ from the ``Millennium Galaxy Catalog'' (Liske et al. 2003) were morphologically analyzed using two component Sersic spheroid\ +\ exponential 
disk decomposition (GIM2D, Simard et al. 2002, 2008). The fraction of pure exponential disks is 14\% and comparable to the fraction of simple disks in our sample. 

The Neighboring Galaxy Catalog (Karachentsev et al. 2004) contains 77 disk-like objects among 451 galaxies with a magnitude limit of ${M_B \ga -12}$ and a distance $\leq$\,10 Mpc. 
This study exhibits an Sd fraction of $21\pm5\%$. This catalog includes low-luminosity galaxies and this increases the completeness of the fraction of simple disks at the faint end. 
Therefore, the simple disk fraction is larger compared to the DR6 sample. 

By comparing the simple disk fractions of this study with its progenitor catalogs, the catalog (Kautsch et al. 2006a) (16\%) and the RFGC (Karachentsev et al. 1999) (17\%), 
also reveal an excellent agreement of the fractions.

This comparison shows a good agreement of the fractions of simple disk galaxies (Sd types) between different studies which are using large numbers of analyzed galaxies. 
The results of this comparison are summarized in Table~\ref{tab3}. Although these studies use different luminosity, distance, inclination and morphological selection
criteria, the average fraction of simple disks is $16.2\% \pm 3.2\%$ among late-type galaxies.


\begin{table}[t]
\caption{
A comparison of the fractions of simple disk galaxies in the present work and other recent studies is shown in this table. The fractions (in percentages) are listed in Col.~1, 
and the method of the morphological classification in Col.~2.}
\label{tab3}
\begin{tabular}{lcl}
\hline\noalign{\smallskip}
 & Fraction [\%] &  Method\\[1.5pt]
\hline\noalign{\smallskip}
This study & 15 & Automated\\
Kautsch et al. 2006a & 16 &Automated\\
Barazza et al. 2008 & 20 & Visual\\
Koda et al. 2007 & 11 &Visual\\
Allen et al. 2006 & 14 &Automated\\
Karachentsev et al. 2004 & 21 & Visual \\
Karachentsev et al. 1999 & 17 & Visual \\[1.5pt]
\hline
\end{tabular}
\end{table}

\section{Summary and discussion}

This study presents the fractions of disk-dominated galaxies and other edge-on disk galaxy types selected from the SDSS DR6. 
Using a quantitative method that measures the disk flatness and the bulge size, 
this sample is divided into morphological classes of galaxies with bulges, intermediate types and apparently bulgeless objects. Each of these broad types contains roughly 1/3 of the 
total sample. Further subdivision is applied to these types and allows us to define the fraction of simple disk galaxies, Sd(f), to be 15\% of the total disk galaxy sample. 
The fractions of other morphological Hubble types of disk galaxies are also presented and compared to the catalog of edge-on disk galaxies (Kautsch et al. 2006a). 

We also present a comparison of the frequency of simple disk galaxies from various galaxy catalogs and find a simple disk fraction of ${\sim\!16\% \pm 3\%}$ on average. 
The fractions are a robust result because of the excellent agreement between the studies,
although various criteria were used to select the galaxies. Small differences of the Sd fractions can be explained qualitatively. 
We conclude that simple disks are a common galaxy type, although formation and evolution models are challenged by explaining bulgeless galaxies. Hence, 
a comprehensive description of simple disks remains to be explored.

 %

%

%

\begin{acknowledgements}
The author expresses his gratitude to the anonymous referee and to Dr. Fabio D. Barazza, Prof. Anthony H. Gonzalez and Leah E. Simon 
for the support of this work and critical reading of the manuscript. 
The author also wants to thank Prof. Eva K. Grebel and Prof. Jay S. Gallagher for their 
kind support. 
Funding for the SDSS ({http://www.sdss.org/}) and SDSS-II has been provided by the 
Alfred P. Sloan Foundation, the Participating Institutions, 
NSF, the U.S. Department of Energy, 
NASA, the Japanese Monbukagakusho, 
and the Max Planck Society, and the Higher Education Funding Council for England. 
The SDSS is managed by the Astrophysical Research Consortium (ARC) 
for the Participating Institutions. The Participating Institutions are the 
American Museum of Natural History, AIP, 
Univ. of Basel, Univ. of Cambridge, Case Western Reserve Univ., 
The Univ. of Chicago, Drexel Univ., Fermilab, 
the Institute for Advanced Study, the Japan Participation Group, 
The Johns Hopkins Univ., the Joint Institute for Nuclear Astrophysics, 
the Kavli Institute for Particle Astrophysics and Cosmology, the Korean Scientist Group, 
the Chinese Academy of Sciences (LAMOST), Los Alamos National Lab., 
MPIA, 
MPA, 
New Mexico State Univ., Ohio State Univ., Univ. of Pittsburgh, 
Univ. of Portsmouth, Princeton Univ., 
USNO, and the Univ. of Washington.
\end{acknowledgements}


\begin{thebibliography}{}
\bibitem{} Abazajian, K., Adelman-McCarthy, J.K., Ag\"{u}eros, M.A., et al.: 2003, AJ 126, 2081
\bibitem{} Abazajian, K., Adelman-McCarthy, J.K., Ag\"{u}eros, M.A., et al.: 2004, AJ 128, 502
\bibitem{} Abazajian, K., Adelman-McCarthy, J.K., Ag\"{u}eros, M.A., et al.: 2005, AJ 129, 1755
\bibitem{} Adelman-McCarthy, J.K., Ag\"{u}eros, M.A., Allam, S.S., et al.: 2006, ApJS 162, 38 
\bibitem{} Adelman-McCarthy, J.K., Ag\"{u}eros, M.A., Allam, S.S., et al.: 2007, ApJS 172, 634
\bibitem{} Adelman-McCarthy, J.K., Ag\"{u}eros, M.A., Allam, S.S., et al.: 2008, ApJS 175, 297
\bibitem{} Allen, P.D., Driver, S.P., Graham, A.W., et al.: 2006, MNRAS 371, 2

\bibitem{} Barazza, F.D., Jogee, S.,  Marinova, I.: 2008, ApJ 675, 1194
\bibitem{} Barnes, J.: 1985, MNRAS 215, 517
\bibitem{} B\"{o}ker, T., Laine, S., van der Marel, R.P., et al.: 2002, AJ 123, 1389
\bibitem{} Cox, T.J.,  Loeb, A.: 2008, MNRAS 386, 461
\bibitem{} Cox, T.J., Jonsson, P., Somerville, R.S., Primack, J.R.,  Dekel, A.: 2008, MNRAS 384, 386
\bibitem{} de Vaucouleurs, G.: 1959, {\it Handbuch der Physik}, Vol. 53,
Astrophysik IV: Stellar Systems, Springer-Verlag, Berlin, G\"{o}ttingen, Heidelberg 
\bibitem{} D'Onghia, E.,  Burkert, A.: 2004, ApJ 612, L13
\bibitem{} D'Onghia, E., Burkert, A., Murante, G.,  Khochfar, S.: 2006, MNRAS 372, 1525

\bibitem{} Fujita, Y.: 2004, PASJ 56, 29
\bibitem{} Goad, J.W.,  Roberts, M.S.: 1979, BAAS 11, 668
\bibitem{} Goad, J.W.,  Roberts, M.S.: 1981, ApJ 250, 79
\bibitem{} Gunn, J.E.,  Gott, J.R., III: 1972, ApJ 176, 1  
\bibitem{} Heidmann, J., Heidmann, N.,  de Vaucouleurs, G.: 1972, MmRAS 75, 85
\bibitem{} Hopkins, P.F., Cox, T.J., Younger, J.D.,  Hernquist, L.: 2008, astro-ph/0806.1739
\bibitem{} Hubble, E.P.: 1936, {\it Realm of the Nebulae},  Yale University
Press, New Haven
\bibitem{} Karachentsev, I.: 1989, AJ 97, 1566
\bibitem{} Karachentsev, I.D., Georgiev, Ts.B., Kajsin, S.S., Kopylov, A.I., et
al.: 1992, A\&AT 2, 265
\bibitem{} Karachentsev, I.D., Karachentseva, V.E.,  Parnovskij, S.L.: 1993, AN 314, 97 (FGC)
\bibitem{} Karachentsev, I.D., Karachentseva, V.E., Kudrya, Yu.N., Sharina, M.E.,  Parnovsky, S.L.: 1999, 
Bull.\ Special Astrophys.\ Obs. 47, 5 (RFGC)
\newpage
\bibitem{} Karachentsev, I.D., Karachentseva, V.E., Huchtmeier, W.K.,  Makarov,
D.I.: 2004, AJ 127, 2031
\bibitem{} Kautsch, S.J., Grebel, E.K.,  Gallagher, J.S., III: 2005, AN 326, 496
\bibitem{} Kautsch, S.J., Grebel, E.K., Barazza, F.D.,  Gallagher, J.S., III: 2005b, AN 326, 595
\bibitem{} Kautsch, S.J., Grebel, E.K., Barazza, F.D., Gallagher, J.S., III: 2006a, A\&A 445, 765 (the catalog)
\bibitem{} Kautsch, S.J., Grebel, E.K., Barazza, F.D.,  Gallagher, J.S., III: 2006b, A\&A 451, 1171
\bibitem{} Kautsch, S.J., Gallagher, J.S., III,   Grebel, E.K.: 2008a,   A\&A, submitted
\bibitem{} Kautsch, S.J., Gonzalez, A.H., Soto, C.A., Tran, K.-V. H., Zaritsky,
D.,  Moustakas, J.: 2008b, ApJ 688, L5 
\bibitem{} Kazantzidis, S., Bullock, J.S., Zentner, A.R., Kravtsov, A.V., 
Moustakas, L.A.: 2008, ApJ 688, 254 
\bibitem{} K\"{o}ckert, F.,  Steinmetz, M.: 2007, in: F. Combes,  J. Palous
(eds.), {\it Galaxy Evolution Across the Hubble Time}, IAU Symp. 235, p.~114
\bibitem{} Koda, J., Milosavljevic, M.,  Shapiro, P.R.: 2007, astro-ph/0711.3014
\bibitem{} Kormendy, J.,  Kennicutt, R.C., Jr.: 2004, ARA\&A 42, 603
\bibitem{} Kormendy, J.,  Fisher, D.B.: 2005, RMxAA 23, 101
\bibitem{} Liske,  J., Lemon,  D.J., Driver,  S.P., Cross,  N.J.G.,  Couch, W.J.: 2003, MNRAS 344, 307
\bibitem{} Mastropietro, C., Moore, B., Mayer, L., et al.: 2005, MNRAS 364, 607
\bibitem{} Matthews, L.D.,  Gallagher, J.S.: 1997, AJ 114, 1899
\bibitem{} Mitronova, S.N., Karachentsev, I.D., Karachentseva, V.E., Jarrett, T.H.,  Kudrya, Yu.N.: 2004, Bull. Special Astrophys. Obs. 57, 5
\bibitem{} Ogorodnikov, K.F.: 1957, SvA 1, 748
\bibitem{} Ogorodnikov, K.F.: 1958, SvA 2, 375
\bibitem{} Okamoto, T., Eke, V.R., Frenk, C.S.,  Jenkins, A.: 2005, MNRAS 363, 1299
\bibitem{} Piontek, F.,  Steinmetz, M.: 2008, in preparation
\bibitem{} Purcell, C.W., Kazantzidis, S., Bullock, J.S.: 2008, astro-ph/0810.2785
\bibitem{} Roediger, E.,  Hensler, G.: 2005, A\&A 433, 875
\bibitem{} Scannapieco, C., Tissera, P.B., White, S.D.M.,  Springel, V.: 2008,
MNRAS 389, 1137
\bibitem{} Simard, L., Willmer, C.N.A., Vogt, N.P., et al.: 2002, ApJS 142, 1 
\bibitem{} Simard, L., Clowe, D.C., Desai, V., et al.: 2008, A\&A, submitted  
\bibitem{} Steinmetz, M.: 2003, Ap\&SS 284, 325
\bibitem{} Stewart, K.R., Bullock, J.S., Wechsler, R.H., Maller, A.H.,  Zentner,
A.R.: 2008, ApJ 683, 597
\bibitem{} Toomre, A.: 1977, in: B.M. Tinsley,  R.B. Larsen (eds.), {\it
Evolution of Galaxies and Stellar Populations},  
	p.~402
\bibitem{} Tran, K.-V.H., Moustakas, J., Gonzalez, A.H., Bai, L., Zaritsky, D., 
Kautsch, S. J.: 2008, ApJ 683, L17 
\bibitem{} Vincent, R.A.,  Ryden, B.S.: 2005, ApJ 623, 137
\bibitem{} Vorontsov-Vel'yaminov, B.: 1967, in: M. Hack (ed.), {\it Modern
Astrophysics},   p.~347
\bibitem{} Vorontsov-Vel'yaminov, B.: 1974, SvA 17, 452
\bibitem{} Walcher, C.J., B\"oker, T., Charlot, S., et al.: 2006, ApJ 649, 692
\bibitem{} York, D.G., Adelman, J., Anderson, J.E., Jr., et al.: 2000, AJ 120, 1579 	

\end{thebibliography}
\end{document}